\documentclass[twocolumn,secnumarabic,amssymb,aps,prb,reprint,floatfix,longbibliography]{revtex4-2}
\usepackage{graphicx}
\usepackage{color}
\usepackage{dcolumn}
\usepackage{bm}
\usepackage{ucs}
\usepackage{physics}
\usepackage{float}
\usepackage{ulem}
\usepackage[T1]{fontenc}
\usepackage{units}
\usepackage{multirow}
\usepackage{booktabs}
\usepackage{placeins}
\usepackage{mathtools}
\usepackage{hyperref}

\begin{document}

\title{Richardson model with complex level structure and spin-orbit coupling for hybrid superconducting islands: stepwise suppression of pairing and magnetic pinning}

\author{Juan Carlos Estrada Salda\~{n}a$^{1}$}
\author{Luka Pave\v{s}i\v{c}$^{3,4}$}
\author{Alexandros Vekris$^{1,2}$}
\author{Kasper Grove-Rasmussen$^{1}$}
\author{Jesper Nyg{\aa}rd$^{1}$}
\author{{Rok \v{Z}itko}$^{3,4}$}

\affiliation{$^{1}$Center for Quantum Devices, Niels Bohr Institute, University of Copenhagen, 2100 Copenhagen, Denmark}
\email{nygard@nbi.ku.dk}
\affiliation{$^{2}$Sino-Danish College (SDC), University of Chinese Academy of Sciences}
\affiliation{$^{3}$Jo\v{z}ef Stefan Institute, Jamova 39, SI-1000 Ljubljana, Slovenia}
\email{rok.zitko@ijs.si}
\affiliation{$^{4}$Faculty of Mathematics and Physics, University of Ljubljana, Jadranska 19, SI-1000 Ljubljana, Slovenia}

\begin{abstract}
An epitaxial semiconductor-superconductor nanowire is a superconducting system with a complex level structure originating from hybridization: in addition to a dense set of higher-energy states derived predominantly from the metallic superconducting shell above the bulk gap $\Delta$, there is a smaller number of lower-energy proximitized states from the semiconducting core that define the induced gap $\Delta^*$. Nanostructures built from such nanowires can furthermore incorporate quantum dots in order to obtain localized spins for storing and manipulating quantum information. We discuss the magnetic field dependence in three devices with different combinations of embedded quantum dots and superconducting islands. For strong fields, they show pinning of excitation energies to a uniform spacing, even if for weak fields they have non-universal properties with different behaviors for even and odd number of confined electrons. We propose a quantum impurity model for hybrid devices that incorporates all relevant physical effects. We show that the model accounts for the key observations and permits unambiguous interpretation in terms of many-particle states. In particular, we study the replicas of the Yu-Shiba-Rusinov states in the hybrid gap, their collapse and oscillation around zero bias with increasing field, and the strong smoothing effect of the spin-orbit coupling (SOC) on these oscillations. We propose that the SOC-induced mixing of many-body states is a generic mechanism and that magnetic pinning is likely to be a ubiquitous feature in hybrid nanowires.
\end{abstract}

\flushbottom
\maketitle
    
\thispagestyle{empty}

\newcommand{\tDe}{\tilde{\Delta}}
\newcommand{\nsc}{n_\mathrm{sc}}
\newcommand{\nscop}{\hat{n}_\mathrm{sc}}

\section{Introduction}

Digital electronics driving modern technology is largely based on semiconducting materials, especially silicon \cite{sze2012,grundman2016}. In recent decades pioneering quantum-coherent solid-state devices were built from superconducting materials \cite{bouchiat1998,nakamura1999,koch2007,blais2020,siddiqi2021,kjaergaard2020,bravyi2022}. Novel functionality can be obtained by combining both types of materials \cite{defranceschi2010,padurariu2010,delange2015,larsen2015,mi2017,benito2020,hays2021,harvey2022,pitavidal2023}. One line of research are semiconducting nanowires covered by a superconducting shell \cite{krogstrup2015epitaxy,kanne2021,pendharkar2021}. They may find application as qubits based on different types of bound states \cite{chtchelkatchev2003,beri2008,aguado2020,flensberg2021,marra2022}. Such semi-super hybrid nanowires have well defined interfaces between the two types of materials, especially if the superconducting shell is deposited epitaxially  \cite{krogstrup2015epitaxy}. This leads to hybridisation of wavefunctions from both sides \cite{mikkelsen2018,winkler2019}, an induced superconducting gap in the semiconducting core (proximity effect) \cite{chang2015Jan} and a non-trivial overall electronic structure, the details of which will need to be well understood for reproducibly building advanced quantum devices \cite{rainis2013,vuik2016,woods2018,Antipov2018Aug,Reeg2018Dec,lai2021}. The local density of states in the centre of the nanowire is dominated close to the edge of the induced gap $\Delta^*$ by the Andreev levels derived from the semiconductor states; this is clearly the case in nanowires in which the proximity gap $\Delta^*$ is significantly lower than the gap of the pristine superconducting material $\Delta$. Using the back gate it is possible to control the relative position of the wavefunctions by moving core states closer or further away from the interface with the shell, which results in variable induced gaps and different magnetic field dependencies \cite{vaitiekenas2018Jul,deMoor2018,bommer2019,vanloo2023}.  The core semiconductor is nominally intrinsic and the devices are typically tuned close to pinch-off points (near depletion), i.e., the free electron density is low. Furthermore, the nanowires have a finite length which leads to quantisation of longitudinal modes, which is expected to be particularly pronounced in the commonly used materials InAs and InSb with very low effective masses of electrons in the conduction band, $m^* \sim 0.04 m_e$ \cite{sanjose2012,cayao2018} (similar quantization effects are also observed in surface adsorbate chains \cite{schneider2021}). The proximitized levels thus do not form a continuum as in a bulk metallic superconductor, but rather a set of discrete states (i.e., a quasicontinuum) \cite{Prada2020Oct,vanloo2023}. In this work, we will present several devices where this is particularly manifest and study the implications of this situation on the magnetic field dependence.

\begin{figure}[t!]
\centering
\includegraphics[width=0.85\linewidth]{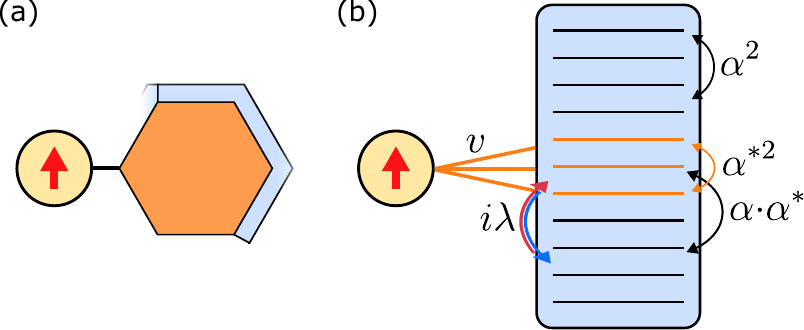}
\caption{
(a) A sketch of a hexagonal semiconducting nanowire (orange) covered with a superconductor (blue), and proximitized with a quantum dot (yellow).
(b) The model consists of a quantum dot coupled to a small number of proximitized levels predominantly located in the semiconducting core, and a large number of strongly superconducting levels in the shell. The superconducting pairing between two levels is a product of their level dependent pairing strength $\alpha$, with $\alpha^* < \alpha$ the weak effective pairing of the proximitized levels. $i \lambda$ describes the all-to-all spin-orbit interaction and $v$ the QD - nanowire coupling strength.
}
\label{sketch}
\end{figure}

For storing quantum information one can confine electrons in quantum dots that can be electrostatically defined in sections of the nanowire. A qubit can be based, for example, on the spin degree of freedom of the trapped particle \cite{heinrich2021}. The resulting local moment is interacting with the itinerant electrons in its vicinity through exchange coupling \cite{anderson1961}. This is known to result in the Yu-Shiba-Rusinov (YSR) states within the superconducting (SC) gap \cite{meden2019anderson,Yu1965,shiba1968classical,rusinov1969theory}. A YSR singlet is a spin-singlet bound state between the spin on the quantum dot (QD) and the spin carried by Bogoliubov quasiparticles. Because the QD wavefunction is localized in the nanowire core, the localized level most strongly couples with the itinerant quasiparticles that are likewise radially confined to the core region, and these are precisely the quasicontinuum proximitized levels introduced above. 
A sketch of the system is shown in Fig.~\ref{sketch}.
In this work, we show that this state of affairs results in a particular behavior in external magnetic field that manifests as magnetic pinning of the sizes of even-occupancy and odd-occupancy regions to equal values (the so-called $1e$-$1e$ charging pattern \cite{Albrecht2016Mar,vanVeen2018Nov,Carrad2020Apr,pendharkar2021}\footnote{
By $1e$-$1e$ charging pattern we mean the absence of even-odd effects in the charging diagram when electrons are added to the system.
}) when the number of electrons in the system is varied. We explain this behavior through a step-by-step collapse of pairing correlation in successive quasicontinuum levels as the field is gradually increased: the linear superposition of zero and double electron occupancy of low-lying proximitized Andreev levels, characteristic of the superconducting state, gives way to singly occupied levels that permit energy lowering through alignment of magnetic moments with the field. This occurs in a step-wise fashion with each change of the state in one additional Andreev level. Each time, the parity of total electron occupancy also changes. Such partial collapse of superconductivity in a discrete set of levels is hence expected to produce oscillatory (zigzag) patterns in charging diagrams plotted as a function of field and gate voltage. In the following we show, however, that spin-orbit coupling (SOC), which is strong in InAs and InSb nanowires  \cite{manchon2015,tosi2019,bommer2019,junger2020}, smoothens out these oscillations and produces a set of equidistant flat lines; this is a direct consequence of the SOC mixing states differing by $\Delta S_z = \pm 1$ that leads to the anticrossing of many-body states. The phenomenon is expected to be robust in the presence of SOC and hence ubiquitous in this family of devices.

This work is structured as follows. In Sec.~\ref{experiment} we present experimental observations on three different devices based on the same batch of hybrid nanowires that indicate the presence of a discrete set of excitations above the edge of the induced superconducting gap (the quasicontinuum) and show the $1e$-$1e$ charging pattern for large magnetic fields. In Sec.~\ref{theory} we propose a simplified theoretical model based on the Richardson's model of a superconductor and the Anderson impurity model of a quantum dot, which incorporates all key ingredients (Cooper pairing, charge repulsion, SOC, Kondo exchange coupling). The Hamiltonian can be written as a product of operator matrices and solved using the density matrix renormalization group (DMRG) in the matrix product state formulation. We show how the successive changes of the electron state in the quasicontinuum levels, in combination with the SOC, lead to generic emergence of the $1e$-$1e$ charging pattern. Finally, in Sec.~\ref{discussion} we discuss the broader implications of this result and conclude by describing how future controlled experiments could be used to conclusively test the theoretical model.

\section{Experimental observations}
\label{experiment}

We start by presenting the evolution of the subgap excitations with increasing magnetic field in three different nanostructures incorporating quantum dots and hybrid SC islands.
For consistent results, the devices are made of the same batch of InAs nanowires with a 7-nm epitaxial superconducting Al shell covering three of their facets. This ensures the same interface (on the average), similar semiconductor/superconductor hybridization, and thus a similar number of low-lying excitations in the hybrid superconductors (HS). The Al is lithographically shaped into the desired pattern by selectively etching the Al using Transene D. We employ standard lock-in techniques to measure the differential conductance $G$ in two-terminal configurations. The measurements are performed in a dilution refrigerator at a base temperature of 25-35 mK. 

\subsection{Evidence for complex level structure}
\label{secquasi}

\begin{figure}[t!]
\centering
\includegraphics[width=0.85\linewidth]{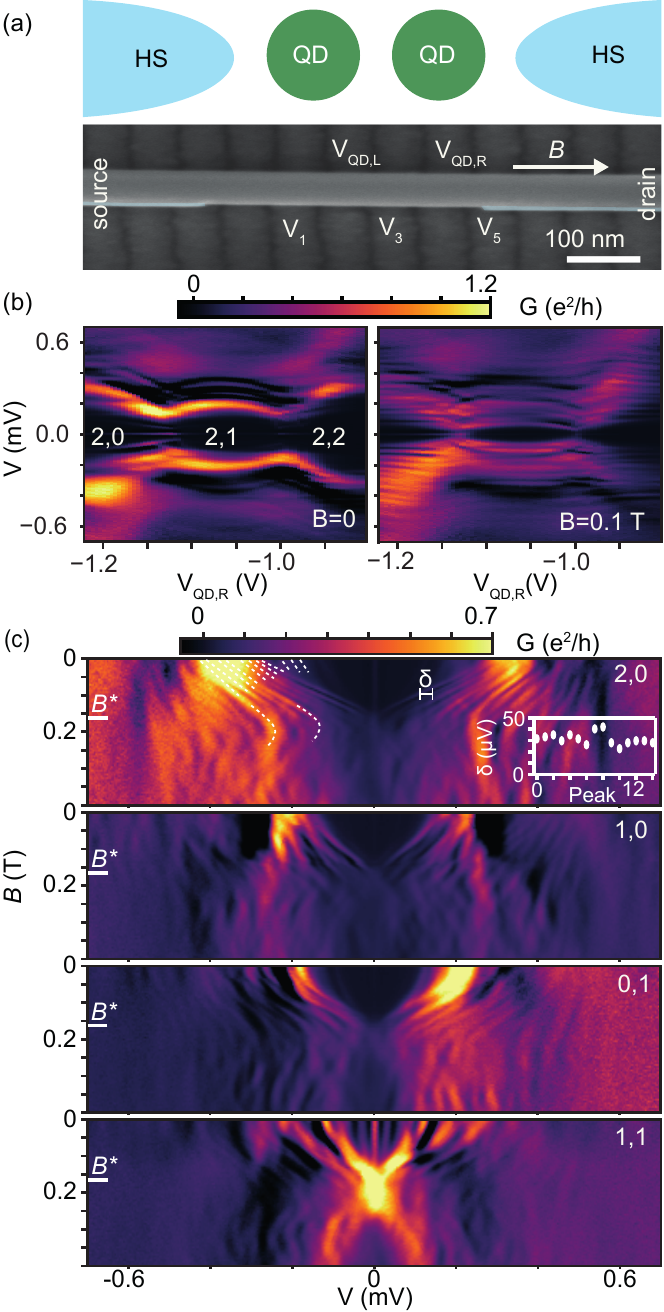}
\caption{(a) Hybrid superconductor--double quantum dot--hybrid superconductor device. Magnetic field direction is indicated by the arrow. (b) Conductance $G$ versus the source-drain voltage $V$ and gate voltage at zero field and at $B=\unit[0.1]{T}$. While only the gate voltage $V_\mathrm{QD,R}$ is indicated, both $V_\mathrm{QD,L}$ and $V_\mathrm{QD,R}$ are swept in a trajectory which changes the ground state by changing the charge R of the right QD while keeping the charge L of the left QD constant. Charge states L,R in the lowest empty QD levels are indicated. (c) Field dependence at four $V_\mathrm{QD,L}$, $V_\mathrm{QD,R}$ locations corresponding to different charge states. The ground state at $B=0$ is a spin singlet in the top and bottom panels, and a spin doublet in the two middle ones. The replica lines converge to $V=0$ at the same $B$ ($B^*=0.16$ T) independently of the charge state. Dashed lines in the top panel highlight the different slopes of some of the replica lines and their curvatures. The inset shows the inter-peak spacing, $\delta=31 \pm 5 \mu$V, at $B=0.1$ T. Average $\delta$ is an estimate for the typical spacing between the Andreev levels that couple to the QD.
}
\label{Fig2}
\end{figure}


We first consider the HS--QD--QD--HS device shown in Fig.~\ref{Fig2}(a). This structure allows high-resolution bias spectroscopy because the sharp excitations of the HS coupled to the first QD can be used to probe the excitation spectrum of the second QD-HS half-system, thereby achieving a very high spectral resolution of 12-14 $\mu$V for the peak width. Fig.~\ref{Fig2}(b) shows the bias spectra for zero field and for $B=0.1$ T. In the absence of the field the spectrum shows a characteristic Yu-Shiba-Rusinov (YSR) loop which appears split due to the way it is probed. In the ideal case of a QD weakly coupled to a standard bulk SC, the exchange interaction would produce at half filling a single excited singlet subgap state with an excitation energy (with respect to the doublet ground state) below the bulk energy gap. In contrast, for our system Fig.~\ref{Fig2}(b) shows that the loop is replicated multiple times below $2\Delta=0.53$ meV, where $\Delta$ is the bulk energy gap of the Al film. We interpret these replicas to be due to multiple discrete Andreev levels in the hybrid superconductor, as commonly observed in similar nanostructures \cite{su2018mirage,estrada2018supercurrent,estradasaldana2020,vanloo2023,steffensen2022}. At finite magnetic field, right panel in Fig.~\ref{Fig2}(b), the number of peaks seems significantly higher than at zero field. This is due to two effects of the Zeeman term, which both enhance our ability to resolve the states forming the quasicontinuum: the energy splitting itself and the decrease of the peak width by a factor of 4 by lifting the degeneracy in the excitations of both QD-HS components.

The field dependence of the differential conductance spectra measured at the center of four different charge sectors, Fig.~\ref{Fig2}(c), shows that the effective mini-gap, defined by the positions of the lowest subgap peaks, closes at the same $B=B^*$ in all cases, i.e., independently of the ground state (GS). This indicates that the phenomenon of gap closure is related to the properties of the nanowire itself (i.e., of the excited states in the two HSs) and that it is effectively independent of the detailed behavior of QDs.

We note that at low fields the spectral lines shift with $B$ with different slopes. This is because the replica lines originate from different orbitals in the HSs that have different gyromagnetic ratios due to mesoscopic fluctuations, as commonly observed in materials with strong SOC \cite{brouwer2000,vonDelft2001,csonka2008,Albrecht2016Mar,vaitiekenas2018Jul,potts2019}. We also observe that some lines are curved. This is due to avoided crossings between the excitations of the same parity, again as a consequence of strong SOC. 
From the inter-peak spacing $\delta$, clearly resolved at $B=0.1$ T, we estimate the number of low-lying excitations in the HSs that are substantially coupled to the quantum dot to be of order 10; for example, the exact count of resolved levels is 12 for the case shown in the top-most panel of Fig.~\ref{Fig2}(c). For $B>B^*$, the zero-bias $G$ never returns to zero, as multiple peaks cross zero bias. For $\delta$ of the order of the peak resolution, the multiple peaks can appear as a single zero-bias peak as in the lowest panel in Fig.~\ref{Fig2}(c).

 \subsection{Evidence for magnetic pinning from the charging diagrams}
 
\begin{figure}
\centering
\includegraphics[width=0.9\linewidth]{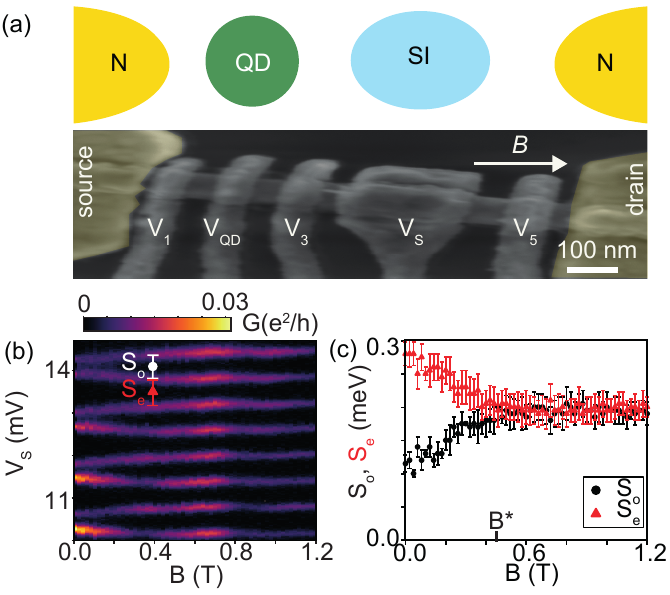}
\caption{(a) Metal--quantum dot--superconducting island (SI)--metal device. (b) Zero-bias $G$ versus $B$ and the SI gate $V_\mathrm{S}$. (c) Spacings $S_\mathrm{o}$, $S_\mathrm{e}$ between adjacent pairs of peaks indicated by bars in (b), as a function of the magnetic field.}
\label{Fig4}
\end{figure}

We now study a metal--QD--superconducting island (SI)--metal device, shown in Fig.~\ref{Fig4}(a), where the QD--SI subsystem is believed to be a clean experimental realization of the Hamiltonian studied in the theory section of this work (Sec.~\ref{theory}). The ground state of this structure is probed by transport measurement using weakly-perturbing tunneling contacts. The QD is tuned so that it has odd occupation, while the gate voltage applied to the SI, $V_S$, is swept so that charge is added to SI in steps of 1, since the SI charging energy exceeds the induced superconducting gap, $E_c > \Delta^*$ (see Ref.~\onlinecite{Saldana2022Jan} for more details on this device structure). In Fig.~\ref{Fig4}(b) we show the zero-bias conductance diagram as a function of magnetic field and $V_S$, with lines of high conductance corresponding to transitions between the different charge states.
With increasing field the widths of even and odd occupancy regions evolves linearly, see also Fig.~\ref{Fig4}(c), until the spacing becomes uniform. The lines are then parallel within the experimental error. This diagram directly shows the $1e$-$1e$ charging pattern for $B>B^*$ \cite{Albrecht2016Mar,Carrad2020Apr,pendharkar2021}.

\subsection{Evidence for magnetic pinning from the large-bias conductance spectra}

\begin{figure}[t!]
\centering
\includegraphics[width=0.9\linewidth]{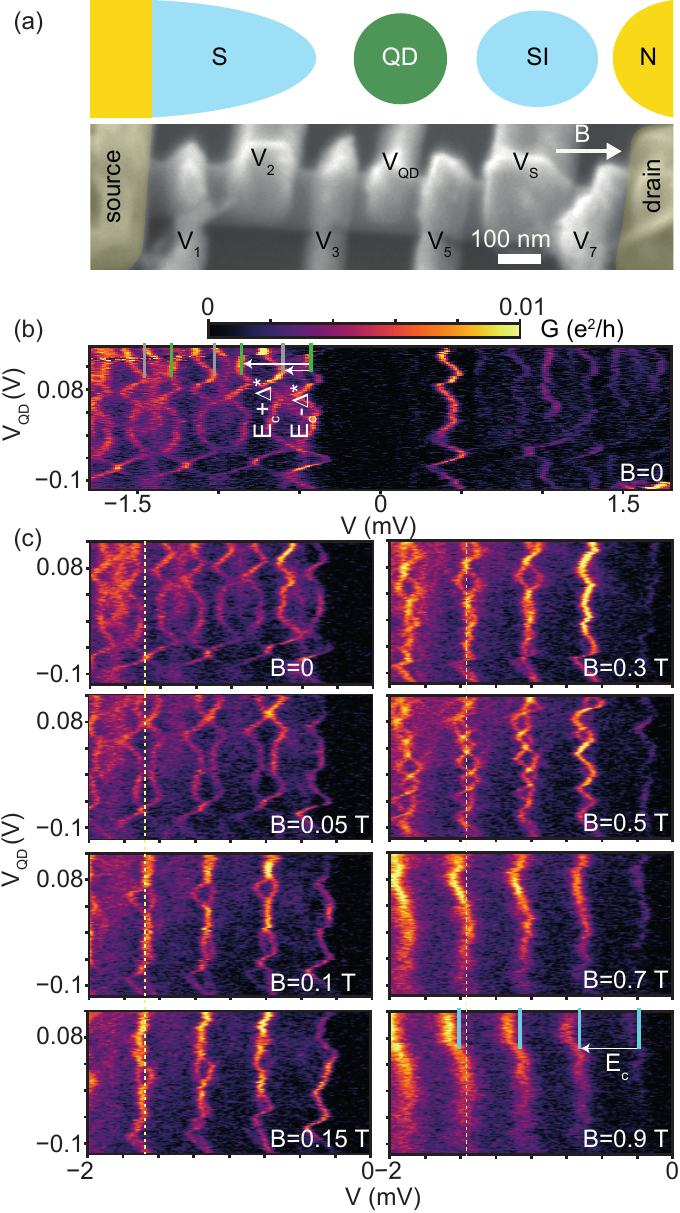}
\caption{(a) Metal--SI--QD--SI--metal device. The left SI is so strongly coupled to the left metal that the device becomes effectively a soft superconductor--QD--SI--metal device. (b) $G$ versus source-drain voltage $V$ and gate voltage $V_\mathrm{QD}$. While only $V_\mathrm{QD}$ is indicated, both $V_\mathrm{QD}$ and $V_\mathrm{S}$ are swept in a trajectory which keeps the total charge approximately constant.
Energy differences for two single-quasiparticle addition excitations of the same total charge are indicated by arrows. The pairs repeat at higher bias with an energy periodicity of $E_\mathrm{c}$, the Coulomb repulsion of the SI. The oscillations of the conductance lines come from filling of the QD and the SI. (c) $B$ dependence of the negative-bias side of (b). A dashed line serves as a guide to the eye for following the convergence of a pair of single quasiparticle addition excitations with increasing $B$. At $B>0.7$ T and up to the largest $B$ in the superconducting state, the conductance lines from each pair are fully merged, and each pair is spaced from the next one by $E_\mathrm{c}$.}
\label{Fig3}
\end{figure}

The pinning can also be probed using an alternative method that makes use of the soft-gap SC--QD--SI--metal device shown in Fig.~\ref{Fig3}(a). The device is characterized in a similar way as in a single-electron transistor, with the bias giving access to higher charge states formed by the coupling of the QD to the SI~\cite{Saldana2022Jan}. The SI is in the regime $E_\mathrm{c}>\Delta^*$, and therefore it can host Cooper pairs but also, for appropriate gate voltage, an odd-numbered quasiparticle. An opaque tunnelling barrier between the QD and the soft-gap superconductor enhances the resolution of single-particle transitions between the higher charge states when probed by its coherence peaks. Fig.~\ref{Fig3}(b) shows a typical bias spectrum at zero field. We sweep the gate voltages of the QD and SI in a trajectory which keeps the total charge approximately constant, therefore maintaining the same type of ground state. This trajectory alternates the even and odd filling of the SI (i.e., absence or presence of a lone quasiparticle) and simultaneously alternates the filling of the QD levels between odd and even numbers of electrons. This results in a modulation of the excitation energy of the state visible as gate-modulated $G$ lines in the spectrum. The lines come in pairs separated by an energy $2\Delta^*$, with the lowest line being unpaired. The pairs repeat themselves at larger bias with every line in a given pair separated by $E_\mathrm{c}$ from its counterpart in the next pair. Therefore, the higher charge states producing the $G$ lines in a given pair have the same charge, which changes in steps of one electron from pair to pair. As shown in Fig.~\ref{Fig3}(c), when applying a $B$ field, the pairs oscillate around each other and steadily collapse into a single line at $B=0.7$ T. They remain collapsed up to $B=1.3$ T, as we corroborated by measurements every 0.1 T. Superconductivity is present in this device up to at least 1.5 T, which excludes the normal state in the superconducting shell as an explanation for the observations. Instead, the emergence of equidistant parallel lines is believed to be due to the same mechanism as the $1e$-$1e$ charging pattern in the zero-bias conductance charging diagrams presented in the previous subsection.

\section{Theoretical modelling}
\label{theory}

The experimental results presented above show two key features: 1) the presence of a quasicontinuum of Andreev levels with energies in the interval $[\Delta^*:\Delta]$, i.e., between the induced and bulk superconducting gap, 2) the $1e$-$1e$ charging pattern for fields $B>B^*$. Using a simplified model we show here how the first feature leads to the second.

\subsection{Model for a superconducting island with complex level structure}
\label{model}

We describe the superconducting island in the hybrid semi-super nanowire using an empirical model based on the Richardson's Hamiltonian for superconductors \cite{vonDelft1996,matveev1997,vonDelft2001}, modified to account for the different types of levels in the system:
\begin{equation}
\begin{split}
H_\mathrm{SI} &= \sum_{i\sigma} \epsilon_i c^\dag_{i\sigma} c_{i\sigma} - \frac{1}{N} \sum_{i,j} \alpha_{i,j} c^\dag_{i\uparrow} c^\dag_{i\downarrow} c_{j\downarrow} c_{j\uparrow} \\
&+ \sum_{i,j,\sigma} \lambda^\sigma_{ij} c^\dag_{i\sigma} c_{j\bar{\sigma}}
+ E_c (\hat{n}_\mathrm{sc}-n_0)^2.
\end{split}
\end{equation}
Here $c_{i\sigma}$ annihilates a spin-$\sigma$ electron in level $i=1,\ldots,N$ with energy $\epsilon_i$ of the SI. For simplicity we take a uniform mesh with $\epsilon_i$ spanning the interval $[-1:1]$; the half-bandwidth thus defines the energy unit in the calculations.

The pairing term takes the form $\alpha_{ij} c^\dag_{i\uparrow} c^\dag_{i\downarrow} c_{j\downarrow} c_{j\uparrow}$ with the coefficient matrix that is a direct product $\alpha_{ij}=\alpha_i \alpha_j$, so that it factorizes:
\begin{equation}
\label{eq2}
\sum_{i,j} \alpha_{i,j} c^\dag_{i\uparrow} c^\dag_{i\downarrow} c_{j\downarrow} c_{j\uparrow} = \left( \sum_i \alpha_i c^\dag_{i\uparrow} c^\dag_{i\downarrow} \right)
\left( \sum_j \alpha_j c_{j\downarrow} c_{j\uparrow} \right).
\end{equation}
The couplings $\alpha_i$ parameterize how strongly a given level $i$ participates in the formation of the superconducting state. Since the levels correspond to linear superpositions of wavefunctions in the superconducting shell and in the semiconducting core, we can separate states according to their character. For states predominantly from the metal shell we assume a constant value, $\alpha_i=\alpha$. The parameter $\alpha$ controls the bulk gap $\Delta$. For states with a strong semiconductor character, we take smaller values, $\alpha_i < \alpha$. Specifically, in numerical calculations we take a set of equally spaced values spanning the interval $[\alpha^*:\alpha]$, such that 
\begin{equation}
    \alpha_i=\alpha^* + (\alpha-\alpha^*)\frac{i-1}{N^*},
\end{equation}
where $i=1,\ldots,N^*$ enumerates the $N^*$ states forming the quasicontinuum of Andreev states with energies below $\Delta$. The parameter $\alpha^*$ controls the proximity gap $\Delta^*$. Reasonable (experimentally relevant) values are $N^* \sim 10$ and $\Delta^*=\Delta/2$; $\alpha$ and $\alpha^*$ need to be tuned appropriately to obtain this gap ratio.

The Hamiltonian also incorporates SOC terms of the form $\lambda^\sigma_{ij} c^\dag_{i\sigma} c_{j\bar{\sigma}}$ with \cite{salinas1999,perenboom1981,halperin1986,vonDelft2001}
\begin{equation}
    \lambda_{ij}^\sigma = -\lambda_{ji}^\sigma = (\lambda_{ji}^{\bar{\sigma}})^*.
\end{equation}
The parameters $\lambda_{ij}$ should be large for the levels derived from the semiconductor states. For simplicity we take 
\begin{equation}
    \lambda_{ij} \equiv i \lambda,
\end{equation}
with $\lambda \in \mathbb{R}$ and $i<j$.
This is a good approximation because the SOC is effective only for singly occupied levels close to the Fermi level, which in our model are precisely those in the semiconducting core.

If the superconducting region is small (an island), the total number of electrons confined in it, $n_\mathrm{sc}=\sum_{i\sigma} c^\dag_{i\sigma} c_{i\sigma}$, is controlled by the charging term $E_c (\nscop-n_0)^2$, where $E_c=e^2/2C$ is the charging energy with $C$ the total capacitance of the SI, and $n_0$ is the gate voltage expressed in dimensionless units signifying the most favorable electron number. 

The quantum dot (QD) in the nanowire is described using the Anderson impurity model:
\begin{equation}
    H_\mathrm{QD} = \sum_\sigma \epsilon n_\sigma + U n_\uparrow n_\downarrow 
     + \sum_{i\sigma} v_i c^\dag_{i\sigma} d_\sigma + \text{H.c.}
\end{equation}
Here $d_\sigma$ annihilates a spin-$\sigma$ electron in the QD level with energy $\epsilon$, while $U$ is the electron-electron repulsion inside the QD; $n_\sigma=d^\dag_\sigma d_\sigma$.
The QD hybridizes mainly with the states in the semiconducting core (i.e., quasicontinuum Andreev levels) because they have a large overlap of their wavefunctions with the QD orbital. We may thus take $v_i \equiv v$ for $i=1,\ldots,N^*$, and zero otherwise.

The Zeeman terms are included both on the QD and on the nanowire levels, pointing either along or perpendicular to the SOC direction:
\begin{equation}
    H_Z = \sum_i g_\mathrm{sc} \mu_B \mathbf{B}\cdot \mathbf{s}_i + g_\mathrm{qd} \mu_B \mathbf{B} \cdot \mathbf{S}_\mathrm{imp}.
\end{equation}
The $g$ factors may be different in nanowire and QD levels and we assume that the magnetic field fully penetrates the superconducting island (i.e., the penetration depth is assumed to exceed the superconductor thickness). For simplicity, in this work we set $g_\mathrm{sc}=g_\mathrm{qd} \equiv g$. When reporting the results of calculation, we will express field strength in terms of the Zeeman energy, $E_Z=g \mu_B B$.
The results reported in the following section will correspond to a magnetic field that is perpendicular to the SOC direction.

\subsection{DMRG solver for complex level structure}

We have extended the implementation of the DMRG solver for problems of coupled QDs and SC islands introduced in Refs.~\onlinecite{Pavesic2021,Saldana2022Jan} to enable numerical studies of the more general Hamiltonian presented above. We were able to construct a compact representation of $H$ in terms of a matrix product operators (MPO) of dimensions $13\times13$.
Such a simple form is possible because of the factorization presented in Eq.~\eqref{eq2}.
The detailed form of the MPO is given in the Appendix. The hybrid nanowire is modelled with $N=100$ levels, $n^*=10$ of which corresponding to proximitized semiconductor states and $90$ to bulk superconductor states. We tuned the bulk coupling $\alpha=0.54$ to set the bulk gap to $\Delta=0.32$, and $\alpha^*$ so that the proximitized gap is half as large, $\Delta^*=0.16$. We set $U/\Delta^*=10$. The charging energy is set to $E_c=0.24=(3/2)\Delta^*$.

In the presence of Coulomb interaction that leads to correlated electron dynamics, the system must be described in terms of many-particle states that are not simple product states of single-particle levels \cite{Pavesic2021,Saldana2022Jan}. A case in point are quantum impurities exhibiting the Kondo effect: the ground state is an entangled state (spin singlet) formed between the impurity local moment and the collective spin degree of freedom in the Kondo cloud of itinerant electrons \cite{yang2017,debertolis2021}. Spectroscopic techniques probe transitions between pairs of many-particle states that in general cannot be reduced to single-particle excitations. This perspective is particularly insightful in quantum impurity systems with superconducting baths: the Yu-Shiba-Rusinov (YSR) subgap peaks correspond to electron-number ($\Delta n=\pm 1$) and spin ($\Delta S_z = \pm 1/2$) changing transitions between even-fermion-parity integer-spin states and odd-fermion-parity half-integer-spin many-particle states \cite{Pavesic2021,Saldana2022Jan}.

\subsection{Level-by-level collapse of pairing correlations in magnetic field}
\label{theory1}

\begin{figure*}[t!]
\centering
\includegraphics[width=1\linewidth]{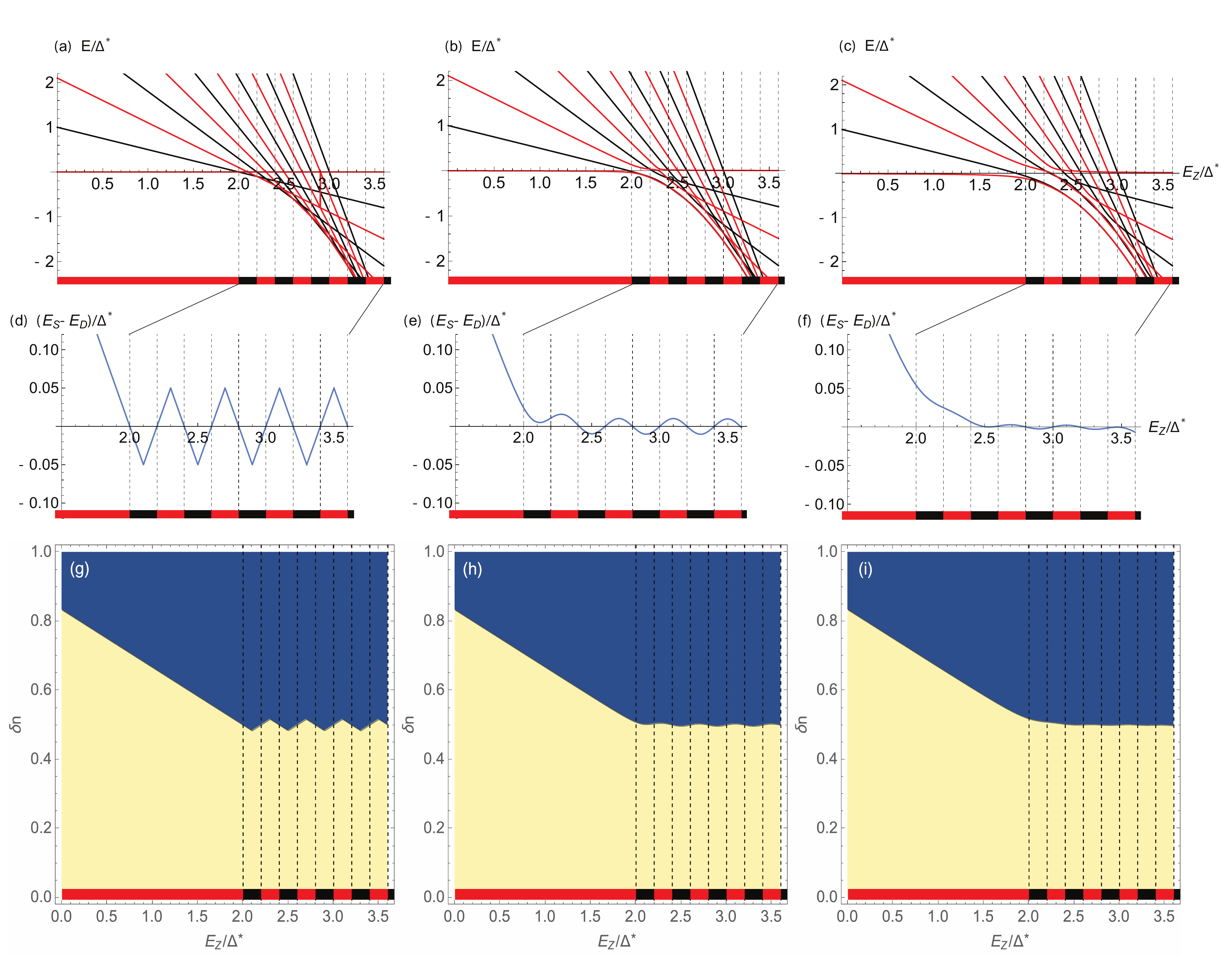}
\caption{Magnetic pinning in nanowires from the perspective of many-particle states. Here we present the results of a toy model calculation. The columns correspond to increasing strength of the SOC with (left to right) $\lambda=0$, $\lambda/\Delta^*=0.075$, $\lambda/\Delta^*=0.15$. (a,b,c) Energies of many-particle eigenstates vs. Zeeman energy; red lines correspond to odd (doublet-like), black lines to even (singlet-like) states; the character of the ground state in a given $E_Z$ interval is indicated through the color of the thick line at the bottom of the plots, the vertical dashed lines indicates the values of $E_Z$ where the ground-state parity switches. Here $\delta n=0.5$.
(d,e,f) Close-up on the energy differences between the lowest integer (even, singlet-like) and lowest half-integer (odd, doublet-like) states. Here $\delta n=0.5$.
(g,h,i) Ground-state diagram in the $(E_Z,\delta n)$ plane. We define $\delta n$ through $n_0=n_\mathrm{ref}+\delta n$ with even integer $n_\mathrm{ref}$.
}
\label{Fig1}
\end{figure*}

Let us first consider the nanowire in isolation; the role of the QD will be investigated later. Level $i$ in the superconductor, assumed to be well separated from other levels (and neglecting the SOC for now), remains in a paired state (i.e., a linear superposition of zero and double level occupancy, $(|0\rangle+|2\rangle)/\sqrt{2}$) for magnetic field strengths $E_Z < 2\tDe_i$, where $\tDe_i$ corresponds to the pairing energy (``effective gap'') of level $i$, with
\begin{equation}
\tilde{\Delta}_i=\frac{\alpha_i}{\alpha} \Delta,
\end{equation}
so that $\tDe_1=\Delta^*$, etc. This follows directly from Eq.~\eqref{eq2} in the limit $N^* \ll N$: the $N-N^* \approx N$ levels generate the bulk gap $\Delta$, while
the $N^*$ low-lying levels experience an effective pairing field reduced by a proportionality factor of $\alpha_i/\alpha$.

For $E_Z$ in excess of $2\tDe_i$, it is energetically favourable for the level $i$ to become singly occupied so that it can reduce its energy by $E_Z/2$ by fully spin polarizing. This can be interpreted as suppression of superconductivity in a single SC level. A key observation is that in terms of many-particle states, this change corresponds to a level crossing between the states with spin polarization $S_z$ and $S_z+1/2$ that have opposite fermionic parity. This implies that there will be a series of such transitions as $E_Z$ increases past $2\Delta^*$, in step-by-step manner for each of the $N^*$ low-lying proximitized states, before the bulk superconductivity is eventually quenched to the normal state at some higher field value. 

This step-by-step process is illustrated in Fig.~\ref{Fig1}(a). The calculation performed here corresponds to a toy model in the $N^* \ll N$ limit without any self-consistency.
We present the many-particle energy spectrum as a function of $E_Z$ for the case of a half-integer value of $n_0$.  For the chosen value of $n_0$, the two relevant states with odd $\nsc$ and even $\nsc$ are those with $\nsc=n_0\pm 1/2$, which have the same charging energy. In this case the energies are  controlled solely by the competition between the pairing interaction and the Zeeman effect. The most noteworthy feature of the plot is the envelope of curves starting at $E_Z=2\Delta^*$, where the collapsing of pairing commences with the level $i=1$ and the ground state $S_z$ starts to increase from 0 in steps of $1/2$ with the increasing magnetic field. The behavior of the low-lying excitations is best represented by the energy difference between the lowest-energy even and the lowest-energy odd many-particle states shown in Fig.~\ref{Fig1}(d). The amplitude of these oscillations is controlled by the gyromagnetic ratio $g$, which sets the slope, and the distance between the level-crossings, which is controlled by the level spacing in the quasicontinuum. 

When $n_0$ is tuned away from a half-integer value, the relative positions of integer and half-integer spin values (as a function of $E_Z$) will shift. Close to even $n_0$, the favored low-lying states are those with even $\nsc$ and integer $S_z$, while close to odd $n_0$ the favored states have odd $\nsc$ and half-integer $S_z$. These energy shifts due to the charging energy terms, combined with the $E_Z$-dependent energy changes seen in Fig.~\ref{Fig1}(d), lead to a zigzag transition line between the odd and even sectors close to half-integer $n_0$ when the state diagram is plotted in the $(E_z,n_0)$ plane, as in Fig.~\ref{Fig1}g. The vertical axis is denoted as $\delta n$, which measures the gate voltage offset from some reference value $n_\mathrm{ref}$ corresponding to an even number of electrons in the ground state of the SI, $n_0=n_\mathrm{ref}+\delta n$.

In the presence of the SOC, the states differing by $\Delta S_z=\pm1$ mix with each other. All level crossings are thus replaced by anti-crossings due to level repulsion. In the toy-model calculation discussed in this subsection, we take this effect into account through off-diagonal matrix elements between the states.  In each charge sector there is then a single low-lying state with continuous evolution as a function of $E_Z$: the ground state in each sector is separated from the excited states with the same fermionic parity by a finite gap whose minimal value is proportional to the SOC strength $\lambda$. The corresponding level diagrams are shown in Fig.~\ref{Fig1}(b),(c) for two values of $\lambda$. Importantly, it can be seen that close to $E_Z \approx 2\Delta^*$, the even (singlet-like) and odd (doublet-like) states come close in energy and then remain very close together, the more so for larger $\lambda$ values. This is even more clearly observed in the energy differences plotted in Fig.~\ref{Fig1}(e),(f): the even-odd oscillations induced by the discrete nature of Andreev levels in the quasicontinuum are progressively washed out. Finally, the level pinning is also revealed in the $(E_Z,n_0)$ diagrams: the zigzag line separating even and odd-parity domains is replaced by a straight line exactly at half-integer values of $n_0$, leading to the $1e$-$1e$ periodicity of the charging diagram in this range of $E_Z$, see Fig.~\ref{Fig1}(h),(i).

\subsection{Role of Yu-Shiba-Rusinov states in magnetic pinning}
\label{theory2}


We now consider the full problem with the QD coupled to the SI. We will show that the general picture established in the previous subsection remains valid in the presence of a QD for high enough magnetic field, irrespective of the QD electron filling (gate voltage tuning) and hybridisation. These two factors only affect the part of the phase diagram in the $(n_0,E_Z)$ space for weak and moderate fields before the superconducting state starts collapsing and the phase diagram reflects the properties of the coupled QD-SI system. For larger fields, when the superconductivity starts collapsing, the phase diagram depends mostly on the physics in the SI region. In other words, the large-field ``magnetic pinning'' behaviour is found to be very universal.

\begin{figure*}[t!]
\centering
\includegraphics[width=1\linewidth]{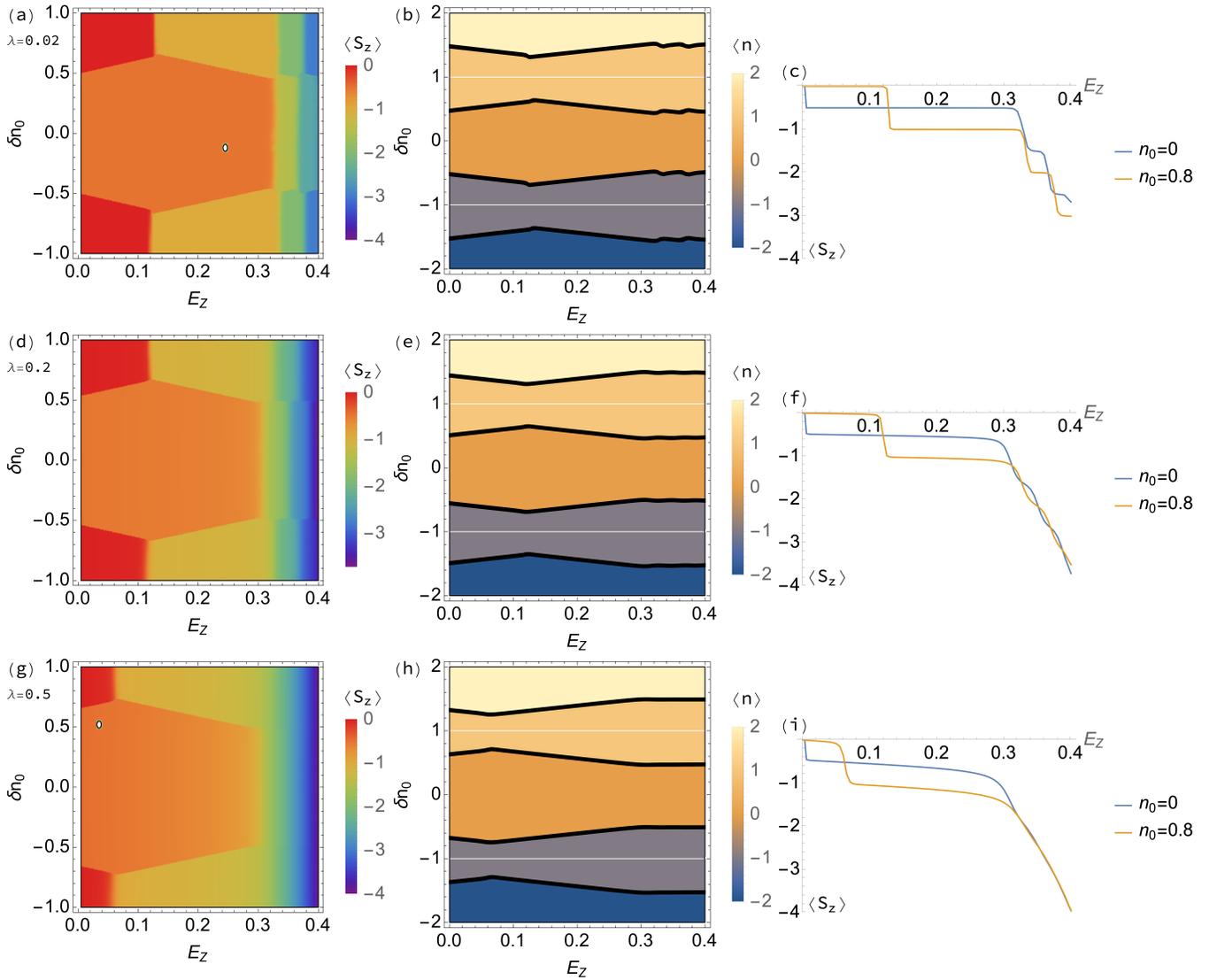}
\caption{Evolution of spin and charge state with increasing field in the SI-QD model. 
(a,d,g) Spin polarization as a function of Zeeman field $E_Z$ and the SI gate voltage $\delta n_0$. (a) Negligible SOC, $\lambda=0.02$. (d) Intermediate SOC, $\lambda=0.2$. (g) Strong SOC, $\lambda=0.5$. The central (orange) region corresponds to the doublet state: for even $\delta n$ and relatively weak hybridisation, the ground state for low $E_Z$ is the ``decoupled spin'' state. This region is surrounded by the singlet state, where odd $\delta n$ tuning provides a quasiparticle that binds antiferromagnetically with the QD spin to form the YSR singlet. At stronger fields, the YSR triplet with ferromagnetic alignment wins over instead. At $E_Z=2\Delta^*=0.32$ the partial collapse of superconductivity commences, leading to a rapid step-wise increase in the spin polarization of the ground state.
(b,e,h) Charge state vs. $E_Z$ and $\delta n_0$. The superimposed black lines correspond to the charge degeneracy lines where enhanced transport through the device is expected. For $E_Z \geq 2\Delta^*$ these lines show $1e$-$1e$ periodicity, and they become increasingly flat, smooth and parallel with increasing SOC strength.
(c,f,i) Line cuts from (a,d,g) at $\delta n_0=0$ and $\delta n_0=0.8$ starting from the doublet and singlet ground state at zero field, respectively. The SOC-induced mixing between the eigenstates with $\Delta S_z=\pm 1$ has multiple effects: 1) the singlet-triplet cross-over moves to lower $E_Z$ and becomes increasingly smooth, 2) the $E_Z<2\Delta^*$ lines become curved, 3) the $E_Z>2\Delta^*$ spin expectation values in even and odd sectors smooth out and practically coincide at high $\lambda$.
(White dots in the figures are numerical artifacts.)
}
\label{figF1}
\end{figure*}

To demonstrate these claims we study an illustrative case of a QD with $U=10\Delta^*$, $E_c/\Delta^*=1.5$, with $N^*=10$ Andreev levels ranging from $\Delta^*$ to the bulk gap of $\Delta=2\Delta^*$. The QD hybridizes with these 10 levels with the hopping amplitude $v=0.09$, which corresponds to the regime of low hybridisation. All energies are again given in units of the half-bandwidth. 
Let us consider first the case of a half-filled quantum dot at weak magnetic fields. Since $U$ is relatively large, the local moment on the QD site is well defined. For even tuning of $\delta n_0$, the ground state is the YSR doublet because of the weak hybridisation. For odd tuning of $\delta n_0$, there is a Bogoliubov quasiparticle sitting in the SI which interacts antiferromagnetically with the QD spin via Kondo exchange coupling to produce the YSR singlet ground state. In the absence of the field, the SOC does not affect the results in any qualitative way, it only slightly shifts the boundary between the doublet and the singlet domains, see panels (a), (d) and (g) in Fig.~\ref{figF1}. With increasing magnetic field the doublet region grows in size due to the energy gain from the Zeeman effect. At a certain value of $E_Z$ the Zeeman energy overcomes the Kondo binding energy in the odd-$\delta n_0$ regions. At this point the spins on the QD and in the SI align: the singlet ground state is replaced by a triplet. In the absence of SOC this is a sharp phase transition (level crossing), see Fig.~\ref{figF1}(a), while in the presence of SOC this is a smooth crossover because the singlet and triplet sectors mix, see Fig.~\ref{figF1}(d,g). Beyond the transition/cross-over point, the doublet region shrinks in size with increasing magnetic field because the energy gain from the Zeeman effect is larger for the triplet than for the doublet. Finally, for $E_Z=2\Delta^*$, we observe the beginning of the gradual collapse of superconductivity in the Andreev levels, with the spin expectation values increasing in steps of 1; this is revealed most clearly in Fig.~\ref{figF1}(c) showing line-cuts at constant $\delta n$. These step-wise changes occur at slightly different values of $E_Z$ for odd and even $n_0$ tuning, giving rise to the zigzag pattern of the phase boundaries, see Fig.~\ref{figF1}(b). With increasing SOC, we observe the phenomenon described in the previous section: the mixing of states differing by $\Delta S_z=\pm 1$ washes out the steps and gives rise to a smooth $E_Z$ dependence, compare panels (c), (f) and (i) in Fig.~\ref{figF1}. Surprisingly, not only are the energy curves for even and odd $n_0$ tuning smooth and parallel, they are in fact nearly overlapping. (One should recall that in the presence of SOC the total spin is not conserved and that the expectation value of its $z$ component is not pinned to half-integer or integer values.) In this regime the total spin remains constant as a function of $n_0$ even if the total electron number is changing in steps of 1. This is the most characteristic feature of the observed pinning phenomenon and explains the near-perfect flatness of the boundary lines for sufficiently large SOC.

\begin{figure*}[t!]
\centering
\includegraphics[width=0.8\linewidth]{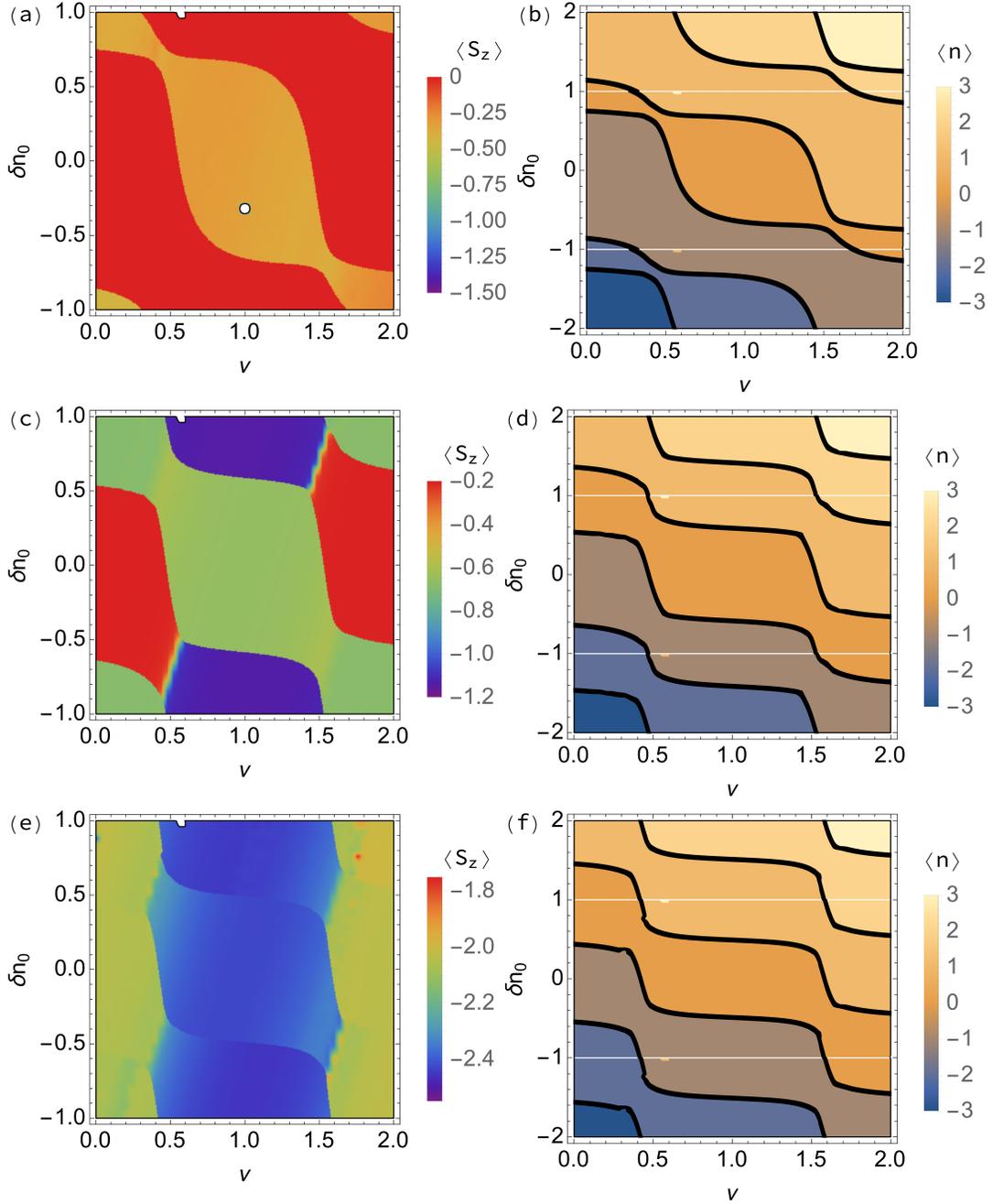}
\caption{Spin and charging diagrams as functions of QD gate voltage $\nu$ and SI gate voltage $\delta n_0$ at strong SOC $\lambda=0.5$ for increasing magnetic field strength. (a,b) Zero field limit, $E_Z=0.001$. (c,d) Intermediate regime, $E_Z=0.2$. (e,f) Partially collapsed SC regime: $E_Z=0.35$.
(a,c,e) Spin polarization. (b,d,f) Charge state. For low field, we observe the conventional charging diagram of the QD-SI device with doublet GS for odd total occupancy and singlet GS for even occupancy. With increasing $E_Z$, the boundary lines become flatter and we observe the singlet-triplet cross-over between even-even and odd-odd gate tuning. For strong field, where $E_Z>2\Delta^*$ we see the consequences of the partial SC collapse: as a function of $\delta n_0$, there is very little variation in spin polarization (as a function of $\nu$, however, one can still switch by a unit of 1/2) and the separation of charge degeneracy lines is $1e$-$1e$.
(Small blotches in the figures are numerical artifacts.)
}
\label{figF2}
\end{figure*}

The robustness of $1e$-$1e$ pinning of the charge regions is clearly illustrated by considering the results in the $(\nu,n_0)$ gate-voltage plane. In Fig.~\ref{figF2} we consider the case of strong SOC for a range of fields. For zero field, Fig.~\ref{figF2}(a,b,c),  we find the typical diagram of even (singlet-like) and odd (doublet-like) regions whose widths (in horizontal and vertical direction) depend on the energy balance, controlled by the charging energies $U$ and $E_c$, the pairing $\Delta$, and the hybridization-dependent binding energy $E_B(v)$. For intermediate fields, Fig.~\ref{figF2}(d,e,f), the diagrams become more complex and include both triplet regions (for the case where both $\nu$ and $n_0$ are odd) and singlet regions (for $\nu$ and $n_0$ both even), as well as the doublet range (for the case where one of $\nu$ and $n_0$ is odd). We note that the boundaries are more parallel (less curved): the increased spin polarization implies decreased electron hopping because of the Pauli exclusion principle. The region widths are still alternating in the vertical direction (i.e., as a function of $n_0$), but less so in the horizontal direction (i.e., as a function of $\nu$). For fields $E_Z>E_Z^*=2\Delta^*$, Fig.~\ref{figF2}(g,h,i), the spacing in the vertical direction becomes uniform, showing the $1e$-$1e$ pattern. In addition, the horizontal width of the regions no longer depends on the electron parity in the superconductor, which can be observed in both the spin and charge stability diagrams. In this regime the QD and the SI are effectively almost decoupled. 

\section{Discussion}
\label{discussion}

The results for the QD-SI system, in particular the occurrence of the $1e$-$1e$ charging pattern for sufficiently large magnetic field, demonstrate the universality of the magnetic pinning mechanism in hybrid nanowires with a quasicontinuum of Andreev levels and a sizeable SOC: the QD parameters and the details of QD-SI coupling only affect how the low-field behavior of the QD connects with the high-field $1e$-$1e$ regime. The spectroscopically observed superconducting ``gap closure'' is thus above all an indication that the pairing correlations in the quasicontinuum states have started collapsing. The SOC leads to a near degeneracy between the lowest-lying even and odd fermion-number-parity states at half-integer $n_0$, which results in a zero-bias anomaly for $B>B^*$. The first finite-energy excitation is another pair of degenerate even and odd states, with the excitation energy controlled by the strength of the SOC. Following the initial gap closure at $B=B^*$, in bias spectroscopy one will thus observe gap reopening for $B>B^*$.

This behavior (zero-bias anomaly, gap reopening, $1e$-$1e$ charging pattern) are consistent with the existence of Majorana zero modes \cite{kitaev2001unpaired} localized at the ends of the hybrid nanowire \cite{lutchyn2010majorana,Oreg2010,Mourik2012May,Shen2018Nov}. Strictly speaking, our model by itself neither supports nor disproves such interpretation, because it is formulated in the orbital/energy space: levels indexed by $i$ are Kramers-pair eigenstates of the non-interacting part of the nanowire Hamiltonian \cite{anderson1959sc} and in this work we make no assumptions about the real-space properties of these levels (such as their localization and spatial extent). There are several commonalities between the models for Majorana zero modes in hybrid nanowires and our Hamiltonian. For example, the partial collapse of superconductivity due to polarization of low-lying Andreev levels can be put in correspondence with the need to obtain effectively spinless superconductors by spin-polarizing electrons with the field in Majorana nanowire models \cite{lutchyn2010majorana,Oreg2010}. Nevertheless, the most that we can infer from our work on this question is that the zero-bias anomaly and the $1e$-$1e$ pattern are generic features of this class of Hamiltonians \cite{lee2014spin} rather than strong signatures of Majorana modes \cite{pan2020}. Additional evidence is necessary for claims of Majorana zero mode existence \cite{frolov2020,flensberg2021}. Since our model is agnostic as regards the properties in real space, there is no requirement for the potential to be smooth; the generality of this Majorana-mimicking behavior is thus greater than that described in Refs.~\onlinecite{kells2012,vuik2019}.

We would like to emphasize the utility of considering this problem from the many-particle perspective. The single-particle approach based on the mean-field picture of superconductivity is not suited for problems where the electron confinement effects are important. In particular, in the presence of Kondo correlations, the single-particle approach is simply not adequate. But even if electron-electron correlations are not particularly strong, the physical interpretation of phenomena can be simpler and more transparent in terms of many-particle states with well defined total charge. The mixing of many-particle states with different magnetic quantum numbers $S_z$ due to SOC is a case in point.

\section{Conclusion}

In this work we explored the magnetic field dependence in several hybrid nanowire devices containing quantum dots and/or superconducting islands. We argued that these devices have a complex spectrum of Andreev levels due to the hybridisation between the states from both sides of the semi-super interface. We devised a quantum impurity model that incorporates such a bath in a simplified way, and showed how Hamiltonians from this family can be solved using the DMRG method. The results uncovered a ubiquitous occurrence of $1e$-$1e$ charging patterns for large magnetic fields in the presence of spin-orbit coupling. We argued that this results from the mixing of many-particle states with different values of magnetic quantum number $S_z$ that results in level repulsion. 

A possible extension of our work, on the side of theory, could go in the direction of higher realism: the coefficients $\epsilon_i$, $\alpha_{ij}$, $\lambda^\sigma_{ij}$ could be derived from realistic real-space models that take a proper account of the crystal properties of materials (including disorder), geometry of the device, potential energy landscape (including the role of gate voltages on metal electrodes), and in particular the detailed properties of the semi-super interface where the states from both sides hybridize. Such realistic model could also serve to produce reference results for controlled experiments designed to more stringently test the hypothesis of partial step-by-step collapse of superconductivity formulated in this work.

\FloatBarrier

\section*{Acknowledgements}

We thank Peter Krogstrup for providing nanowire materials. The project received funding from the European Union’s Horizon 2020 research and innovation program under the Marie Sklodowska-Curie grant agreement No.~832645. We additionally acknowledge financial support from the Carlsberg Foundation, the Novo Nordisk Foundation SolidQ project, the Independent Research Fund Denmark, QuantERA ’SuperTop’ (NN 127900), the Danish National Research Foundation, Villum Foundation project No.~25310, and the Sino-Danish Center. L.~P. and R.~\v{Z}. acknowledge the support from the Slovenian Research Agency (ARRS) under Grants No.~P1-0044, P1-0416 and J1-3008.

\appendix

\begin{widetext}
\section{Matrix product operator representation}

The matrix product representation of the Hamiltonian from Sec.~\ref{theory} is $H=\prod_{i=0}^N W_i$.
The first element is a vector and defines the site with the quantum dot.
\begin{equation}
W_0 = \begin{pmatrix}
I &
\Big[ \epsilon \hat{n}_\mathrm{QD} + \frac{E_Z}{2} (\hat{n}_{\mathrm{QD}\uparrow} - \hat{n}_{\mathrm{QD}\downarrow}) + \frac{E_X}{2} (S^+ + S^-) + U \hat{n}_{\mathrm{QD}\uparrow} \hat{n}_{\mathrm{QD}\downarrow}  \Big] &
- d_{\uparrow} F_1 &
- d_{\downarrow} F_1 &
  d^\dagger_{\uparrow} F_1 &
  d^\dagger_{\downarrow} F_1 &
    0   &
    0   &
    0   &
    0   &
    0   &
    0   &
    0
\end{pmatrix}.
\end{equation}
$d_\sigma$ are the quantum dot operators and $\hat{n}_{\mathrm{QD},\sigma} = d^\dagger_\sigma d_\sigma$ and  $\hat{n}_\mathrm{QD} = \hat{n}_{\mathrm{QD},\uparrow} + \hat{n}_{\mathrm{QD},\downarrow}$.
$F_j = (-1)^{\hat{n}_j}$ is the local fermionic parity operator. It takes care of the fermionic anticommutation rules by giving a factor of $-1$ if the $j$-th level is occupied by a single particle.
$E_Z$ is the Zeeman splitting due to the magnetic field in the $z$ direction and $E_X$ in the $x$ direction, with $S^+ = d^\dagger_\uparrow d_\downarrow$ and $S^- = d^\dagger_\downarrow d_\uparrow$ the spin raising and lowering operators. The SOC direction is along the $x$ axis, thus $E_Z$ corresponds to a perpendicular, and $E_X$ to parallel magnetic field.

A general site $j$ of the system is represented by a matrix:
\begin{equation}
W_j = \begin{pmatrix}
1 & X_j & 0 & 0 & 0 & 0 & 0 & 0 & 0 & 
c_{j\downarrow} & c_{j\uparrow} & c^\dagger_{j\downarrow} & c^\dagger_{j\uparrow}  \\
0 & I & 0 & 0 & 0 & 0 & 0 & 0 & 0 & 0 & 0 & 0 & 0 \\
0 & v_j c^\dagger_{j\uparrow} & F_j & 0 & 0 & 0 & 0 & 0 & 0 & 0 & 0 & 0 & 0 \\
0 & v_j c^\dagger_{j\downarrow} & 0 & F_j & 0 & 0 & 0 & 0 & 0 & 0 & 0 & 0 & 0 \\
0 & v_j c_{j\uparrow} & 0 & 0 & F_j & 0 & 0 & 0 & 0 & 0 & 0 & 0 & 0 \\
0 & v_j c_{j\downarrow} & 0 & 0 & 0 & F_j & 0 & 0 & 0 & 0 & 0 & 0 & 0 \\
0 & y_j c^\dagger_{j\uparrow}c^\dagger_{j\downarrow} & 0 & 0 & 0 & 0 & 1 & 0 & 0 & 0 & 0 & 0 & 0 \\
0 & y_j c_{j\downarrow}c_{j\uparrow} & 0 & 0 & 0 & 0 & 0 & 1 & 0 & 0 & 0 & 0 & 0 \\
0 & \hat{n}_j & 0 & 0 & 0 & 0 & 0 & 0 & 1 & 0 & 0 & 0 & 0 \\
0 & i \lambda c^\dagger_{j\uparrow} & 0 & 0 & 0 & 0 & 0 & 0 & 0 & F_j & 0 & 0 & 0 \\
0 & i \lambda c^\dagger_{j\downarrow} & 0 & 0 & 0 & 0 & 0 & 0 & 0 & 0 & F_j & 0 & 0 \\
0 & i \lambda c_{j\uparrow} & 0 & 0 & 0 & 0 & 0 & 0 & 0 & 0 & 0 & F_j & 0 \\
0 & i \lambda c_{j\downarrow} & 0 & 0 & 0 & 0 & 0 & 0 & 0 & 0 & 0 & 0 & F_j 
\end{pmatrix},
\end{equation}
where the on-site potential term for the $j$-th level is
\begin{equation}
    X_j = [\epsilon_j + E_c (1-2 n_0)] \hat{n}_j + \frac{E_Z}{2} \left(\hat{n}_{j\uparrow} - \hat{n}_{j\downarrow} \right) + \frac{E_X}{2} \left( S_j^+ + S_j^- \right) + (y_j^2 g + 2 E_c) \hat{n}_{j\uparrow} \hat{n}_{j\downarrow}.
\end{equation}
$g = 2\alpha/N$ is a measure of the pairing potential with $y_j$ its modulation, so that $\alpha_{ij} = y_i y_j \alpha$.
The spin orbit coupling is $\lambda$. To account for level-dependent SOC, a similar procedure as with the modulation of pairing can be implemented.

Finally, the last, $N$-th, superconducting level is represented by a vector:
\begin{equation}
W_N = \begin{pmatrix}
X_N &
 1 &
 v_N c^\dagger_{N \uparrow}  &
 v_N c^\dagger_{N \downarrow}&
 v_N c_{N \uparrow}  &
 v_N c_{N \downarrow}  &
 c^\dagger_{N \uparrow} c^\dagger_{N \downarrow}   &
 c_{N \downarrow} c_{N \uparrow}   &
 \hat{n}_N   &
    i \lambda c^\dagger_{N \uparrow}   &
    i \lambda c^\dagger_{N \downarrow} &
    i \lambda c_{N \uparrow}   &
    i \lambda c_{N \downarrow}
\end{pmatrix}.
\end{equation}

\end{widetext}

\bibliography{sample}

\end{document}